\documentclass[aps,reprint,amsmath,amssymb,superscriptaddress]{revtex4-2}

\usepackage{amsmath}    
\usepackage{amssymb}
\usepackage{graphicx}   
\usepackage{verbatim}   
\usepackage[colorlinks,urlcolor=blue,citecolor=blue,linkcolor=blue]{hyperref}
\usepackage{color}      
\usepackage{subfigure}  
\usepackage{hyperref}   
\usepackage{dcolumn}
\usepackage{bm}
\usepackage{epsfig}
\usepackage{epstopdf}
\usepackage{braket}
\usepackage{bbold}
\usepackage{xcolor}

\begin{document}

\title{Ultrafast Single-Qubit Gates in the Diabatic Regime}

\author{Deniz T\"{u}rkpen\c{c}e}
\affiliation{\.{I}stanbul Technical University, Informatics Institute, 34469 Maslak, \.{I}stanbul, T\"{u}rkiye}
\affiliation{Qready Quantum Technologies, ITU ARI Teknokent, 34467 \.{I}stanbul, T\"{u}rkiye}

\author{Sel\c{c}uk \c{C}akmak}
\affiliation{Department of Software Engineering, Samsun University, 55420 Samsun, T\"{u}rkiye}

\email[]{selcuk.cakmak@samsun.edu.tr}

\date{\today}

\begin{abstract}
Logic gates in superconducting quantum processors are implemented through precise quantum control techniques in the microwave regime. The choice of drive frequency and other control parameters directly determines the duration of quantum gate operations. Because current devices remain too noisy to reach fault tolerance, reducing gate durations, and thereby the overall circuit depth, is of critical importance. In this work, we present a model of single qubit gate execution in both the adiabatic regime, where the rotating wave approximation (RWA) is valid, and the diabatic regime, where the RWA no longer applies. Using parameters representative of superconducting qubits, we investigate how gates can be driven at durations well below conventional timescales, and we examine the associated limitations, performance trade offs. The results demonstrate that ultrashort control pulses in the diabatic regime can achieve fidelities comparable to those obtained under standard RWA conditions, offering a possible route to faster quantum logic without sacrificing accuracy under idealized conditions.
\end{abstract}
\maketitle
\section{Introduction}
Current efforts to build quantum computers are advancing at a tremendous pace. However, current quantum computers remain in the Noise Intermediate-Scale Quantum (NISQ) era~\cite{preskill_quantum_2018}. Among various implementations, superconducting platforms, particularly those employing transmon qubits, are at the forefront in terms of both scalability and qubit control maturity~\cite{koch_charge-insensitive_2007, krantz_quantum_2019}. 

According to the current state-of-the-art achieved through intensive work in this field, fidelity values of 0.999 have been reached for single-qubit logic gates and 0.995 for two-qubit logic gates in superconducting quantum computers~\cite{bluvstein_logical_2024, kim_evidence_2023}. The fidelities achieved for single-qubit logic gates, in particular, can be considered sufficient and efficient enough not to require further improvement. 
Nevertheless, significant limitations exist in circuit depth when considering the gate counts or implementation times of both single and multi-qubit logic gates~\cite{kim_evidence_2023,arute_quantum_2019}. These limitations are considered the primary obstacles to applying quantum computers to real-world problems. In addition to ongoing efforts to suppress noise effects, the pursuit of shorter-duration quantum gate implementations is also being actively explored in parallel, with the aim of building more efficient quantum computers~\cite{werninghaus_leakage_2021,negirneac_high-fidelity_2021}.

The implementation of circuit-level quantum logic gates requires a time-dependent, pulse-level analysis of the interaction between control fields and superconducting artificial atoms~\cite{danin_procedure_2024,alexander_qiskit_2020,li_pulse-level_2022,earnest_pulse-efficient_2021}. At the device level, the implementation of a single-qubit gate can be described within the framework of the RWA, addressing a fundamental scenario in quantum optics: a two-level atom (qubit) driven by a classical field~\cite{shirley_solution_1965,scully_quantum_nodate, mandel_optical_2008}. The RWA is a weak-coupling approximation that neglects counter-rotating terms under near-resonance conditions. It breaks down under strong driving fields or when the field parameters undergo sudden changes~\cite{shore_simple_1990}.

Previous studies~\cite{kaplan_time_2004,shakov_sudden_2006} have addressed sudden changes in the driving field and identified conditions under which pulse modulation parameters can be analyzed in the diabatic (or `kicked') regime. In our work, we revisit similar calculations motivated by the goal of creating ultra-short-duration logic gates in the microwave domain – the operational range of transmon qubits – and present the necessary conditions and limitations
\section{Preliminaries}
We consider a two-level system interacting with a time-dependent classical drive field, described by the Hamiltonian
\begin{equation}\label{Eq:Hamilton}
H(t) = \frac{\hbar\omega_0}{2}\sigma_z + \Omega(t)\cos(\omega_D t) \sigma_x,
\end{equation}
where $\hbar$ is the reduced Planck constant, $\omega_0$ is the transition frequency between the eigenstates of the two-level atom to be controlled, and $\omega_D$ is the frequency of the classical drive field. The operators $\sigma_z$ and $\sigma_x$ are the usual Pauli matrices. Here, $\Omega(t) = \mu E(t)/\hbar$ represents the time-dependent Rabi frequency, with $\mu$ denoting the transition dipole moment and $E(t)$ the electric field envelope of the classical drive.

The temporal modulation of the classical electric field is
\begin{equation}\label{Eq:Gauss}
E(t) = A_0 \exp\left(-\frac{(t-t_0)^2}{\tau^2}\right)
\end{equation}
chosen as a Gaussian envelope, which is a commonly adopted form in quantum control methods~\cite{danin_procedure_2024,alexander_qiskit_2020,kuzmanovic_high-fidelity_2024}. 
Here, $A_0$ is the peak amplitude of the electric field, $t_0$ is the pulse center time, and $\tau$ is the pulse duration. In addition, we have set the amplitude of the electric field to be $\alpha$ by $A_0 = \frac{\alpha\hbar}{\mu\sqrt{\pi}\tau}$, consistent with the pulse area theorem~\cite{allen_optical_1987, mccall_self-induced_1969} valid within the Rotating Wave Approximation (RWA) (see Appendix~\ref{appendix:pulse}).

Due to the reasons outlined in the introduction, the implementation of quantum gates should, in general, be described by open-system dynamics that account for noise and decoherence. However, in our case, we focus on scenarios involving isolated, single-pulse operations in the diabatic regime—characterized by extremely short pulse durations. Given these specific conditions, we model the system evolution using time-dependent Schrödinger equation for a two-level system is given by $i\hbar \frac{d}{dt}|\psi(t)\rangle = H|\psi(t)\rangle$ as 

\begin{equation}\label{Eq:Schrödinger}
i\hbar \frac{d}{dt} \begin{pmatrix} a_0(t) \\ a_1(t) \end{pmatrix} = \begin{pmatrix} \frac{\hbar\omega_0}{2} & \tilde{\Omega}(t) \\ \tilde{\Omega}(t) & -\frac{\hbar\omega_0}{2} \end{pmatrix} \begin{pmatrix} a_0(t) \\ a_1(t) \end{pmatrix}
\end{equation}
where $\tilde{\Omega}(t)=\Omega(t)\cos(\omega_D t)$. Here, $|\psi(t)\rangle = a_0(t)|0\rangle + a_1(t)|1\rangle$, where $|0\rangle$ and $|1\rangle$ are the computational basis states, and $a_0(t)$, $a_1(t)$ are the complex coefficients satisfying $|a_0(t)|^2 + |a_1(t)|^2 = 1$.

By the solution to Eq.~(\ref{Eq:Schrödinger}), the time evolution of the state vector is given by $|\psi(t)\rangle =U(t)|\psi(0)\rangle$, where $U(t)$ is the time-evolution operator defined as:
\begin{equation}\label{Eq:Ordering}
U(t) = \mathcal{T} \exp\left(-\frac{i}{\hbar} \int_{0}^{t} H(t') dt'\right)
\end{equation}
Here, $\mathcal{T}$ denotes the time-ordering operator, which ensures that operators are applied in chronological order, from earlier to later times. 

\subsection{System Dynamics Across Control Regimes}

The behavior of the two-level system under control field modulation is critically dependent on the relationship between the pulse duration and the system's intrinsic timescales. When the Rabi frequency changes slowly relative to the qubit's energy scale—specifically, when the pulse duration $\tau$ is much longer than the qubit's inverse transition frequency ($\omega_0 \tau \gg 1$)—the system is considered to be in the adiabatic regime~\cite{shore_simple_1990}. In this limit, the Hamiltonian $H(t)$ at different times approximately commutes, and consequently, the effects of the time-ordering operator $\mathcal{T}$ become negligible. As a result, the system's state closely follows the instantaneous eigenstates of $H(t)$, allowing for smooth transitions between eigenstates driven by the modulated control field $\tilde{\Omega}(t)=\Omega(t)\cos(\omega_D t)$.
In this regime, RWA often provides a highly accurate description and successfully models the system's dynamics.

Conversely, when rapid changes occur in the driving field, such that the pulse duration $\tau$ is comparable to or much shorter than the qubit's intrinsic timescale ($\omega_0 \tau \ll 1$), the system enters the diabatic (or `kicked') regime~\cite{kaplan_time_2004}. In this regime, the Hamiltonian $H(t)$ changes rapidly, and its values at different times generally do not commute, i.e., $[H(t_1), H(t_2)] \neq 0$ for $t_1 \neq t_2$. Therefore, the time-ordering operator $\mathcal{T}$ in the evolution operator $U(t)$ becomes crucial for accurately describing the system's dynamics. For extremely short durations, the pulse shape can be approximated as a Dirac delta function, $E(t) \approx \delta(t-t_0)$, exhibiting a sharp peak around $t_0$. It is important to emphasize that in this regime, the RWA breaks down, a consequence of both the non-commuting nature of the Hamiltonian at different times and the strong field-system interaction.

\subsection{Kicked Evaluation}
By the formal integration of Eq.~(\ref{Eq:Ordering}), the time evolution operator in the kicked regime can be represented as~\cite{kaplan_time_2004,shakov_sudden_2006, altintas_entanglement_2011, altintas_control_2013}
\begin{equation}
U^{k}(t) = 
\begin{pmatrix}
e^{i\omega_0 t} \cos\alpha & -i e^{i\omega_0(t - 2t_0)} \sin\alpha \\
-i e^{-i\omega_0(t - 2t_0)} \sin\alpha & e^{-i\omega_0 t} \cos\alpha
\end{pmatrix}.
\end{equation}
Although the time evolution of a two-level quantum system is always described by a unitary operator in the SU(2) group, the specific form of this operator depends on the structure of the driving field. In the case of a constant Hamiltonian such as $H = \frac{\omega}{2} \sigma_x$, the evolution is a rotation around the x-axis: 
$
R_x(\alpha) = \exp\left(-i \frac{\alpha}{2} \sigma_x \right).
$
However, in the presence of a time-dependent drive, especially in the diabatic regime, the resulting evolution remains unitary and within SU(2), but it no longer corresponds to a pure x-rotation. Instead, it takes the more general form
$
U(t) = \exp\left(-i \alpha \, \hat{n}(t) \cdot \vec{\sigma} \right),
$
where \( \hat{n}(t) \) is a time-dependent rotation axis. In this regime, counter-rotating terms contribute significantly to the dynamics, making the evolution deviate from the simple RWA-based behavior.

This work primarily concentrates on the realization of single-qubit logic gates, with a specific emphasis on the NOT gate. Therefore, crucial quantifiers in our computational analysis include the expectation value $\langle\sigma_z(t)\rangle = |a_0(t)|^2 - |a_1(t)|^2$ and the Bloch vector trajectories, which collectively facilitate a clear observation of the state transition between computational basis states $|0\rangle$ and $|1\rangle$ (and vice versa). From a physical implementation perspective, a proper NOT gate should ideally induce a transition from one basis state to the other without leaving any residual superposition. To assess the fidelity of this implementation in more detail, it is imperative to monitor whether coherence is unintentionally generated during the pulse application. To this end, the $l_1$ norm of coherence, formally expressed as $C_{l_1}(t) = 2 \left| a_0(t) \right| \left| a_1(t) \right|$ for a single qubit system, serves as an essential metric for gauging the performance of the physical implementation.

\begin{figure*}
\includegraphics[width=7.0in]{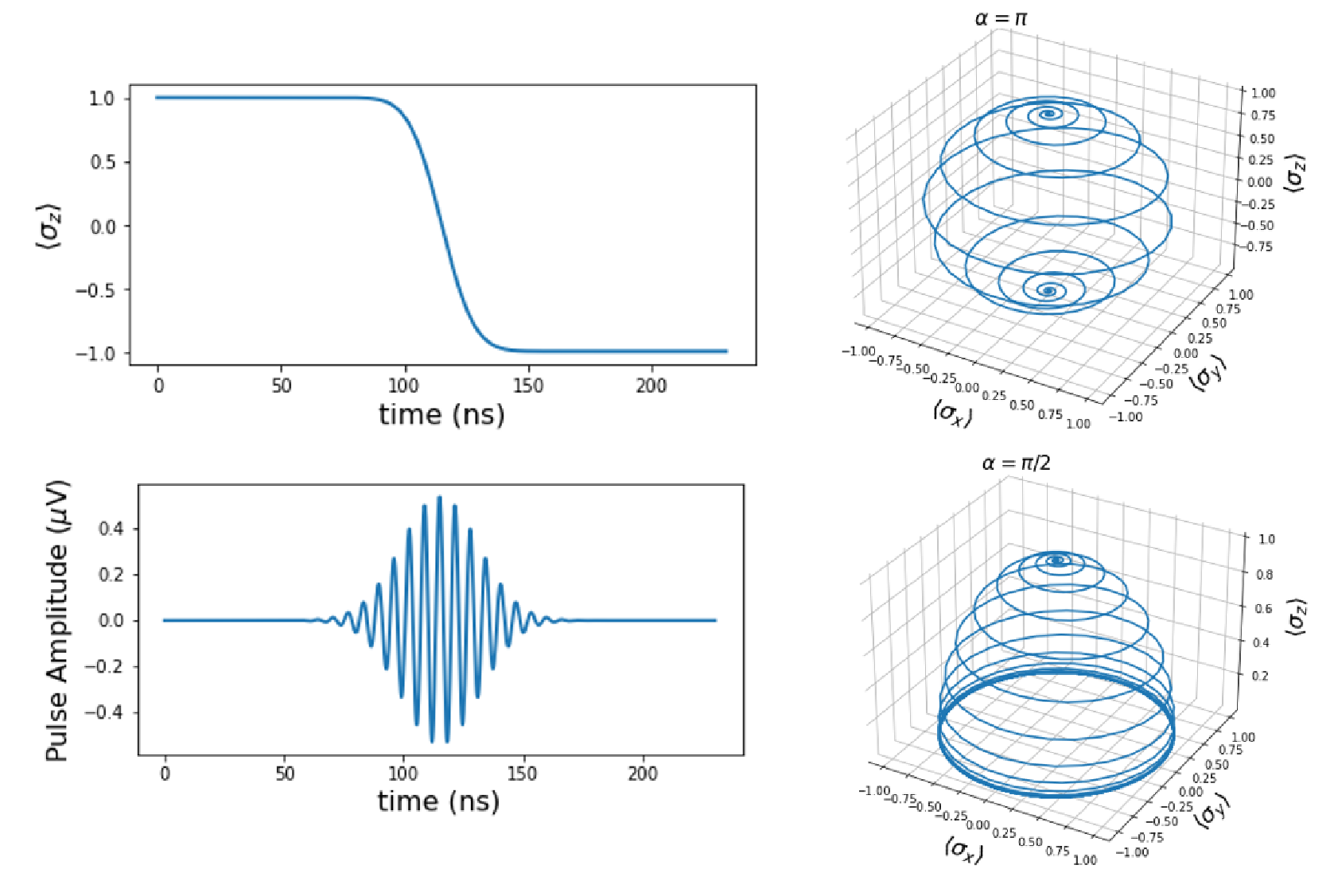}
\caption{ \label{fig:bloch} (Colour online.)Overview of the qubit dynamics under a Gaussian-modulated classical drive. 
(Upper-left panel) Time evolution of the qubit magnetization $\langle\sigma_z(t)\rangle$ for an initial state $|0\rangle$, indicating a population transfer to the state $|1\rangle$. 
(Bottom-left panel) The Gaussian-modulated classical drive pulse, with a carrier frequency $\omega_D = 2\pi \times 4.5~\mathrm{GHz}$, pulse amplitude is $\sim$54$\mu$V and the pulse duration is $\tau = 23~\mathrm{ns}$. (Upper-right panel) Bloch vector trajectory during evolution under a $\pi$-pulse ($\alpha = \pi$).
(Lower-right panel) Bloch vector trajectory for a $\pi/2$-pulse ($\alpha = \pi/2$), showing a rotation from the north pole to the equatorial plane of the Bloch sphere. } 
\end{figure*}

\section{Gate dynamics: RWA and Beyond}
In this section, we investigate the dynamics of a single qubit under classical driving, comparing the kicked regime with the adiabatic regime where the rotating wave approximation (RWA) holds. As a typical realization of a NOT gate applied to a single qubit using quantum control methods in the microwave regime, we implement the dynamics governed by Eq.~\ref{Eq:Hamilton}. In our numerical simulations, we adopt natural units where $\hbar = 1$.

Figure~\ref{fig:bloch} illustrates the time evolution of the qubit under resonant classical driving. The upper-left panel of Fig.~\ref{fig:bloch} displays the evolution of the qubit magnetization $\langle \sigma_z(t) \rangle$, showing that the qubit, initially prepared in the state $|0\rangle$, is successfully driven to the $|1\rangle$ state. The driving field is applied with a Gaussian envelope, a commonly accepted modulation scheme in practical implementations, as defined in Eq.~\ref{Eq:Gauss}. The lower-left panel shows the pulse shape applied over a timescale on the order of nanoseconds. The carrier frequency is chosen to match the qubit’s transition frequency $\omega_0 = 2\pi \times 4.5$ GHz, ensuring resonant driving. The dipole moment value has been chosen as $\mu = 3 \times 10^{-25}~\mathrm{C \cdot m}$, which corresponds to a realistic range of $0.1$–$1\,e\text{\AA}$ reported for superconducting qubits~\cite{lisenfeld_electric_2019}.

By substituting the relevant parameters into the expression for $A_0$ and setting $\alpha = \pi$, the peak value of the Rabi frequency is calculated as $\Omega/2\pi \approx 12.3$ MHz. Since $\Omega \ll \omega_0$, the system dynamics can be reliably analyzed within the RWA framework. For a detailed discussion on converting the peak electric field amplitude into a voltage amplitude, see Appendix~\ref{appendix:Voltage}.

\begin{figure*}
\includegraphics[width=7.3in]{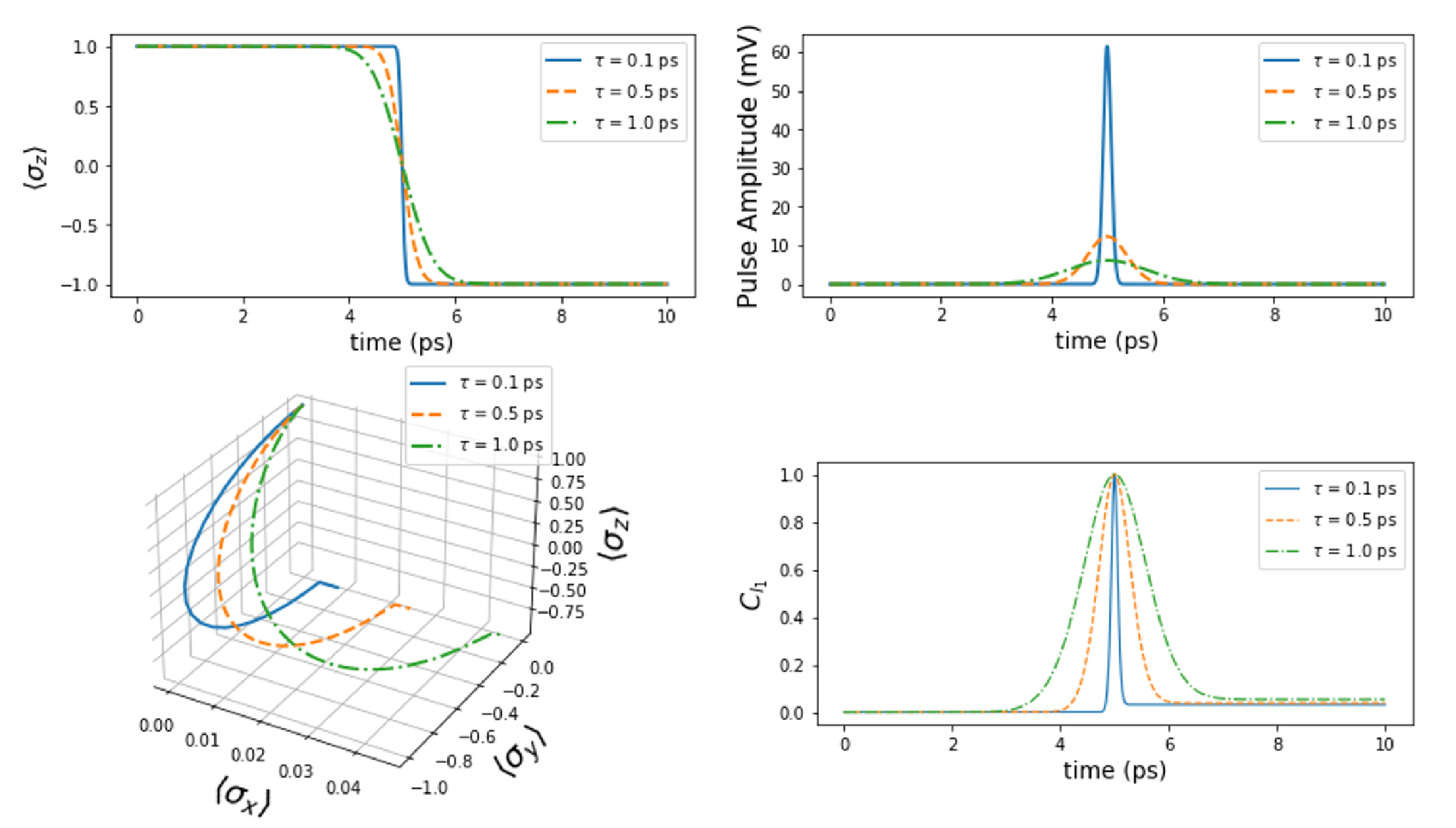}
\caption{ \label{fig:Pico} (Colour online.) Overview of the qubit dynamics under a kicked classical drive. The dynamics is examined for three different pulse durations: $\tau = 0.1$ ps, $\tau = 0.5$ ps, and $\tau = 1$ ps.  
(Upper-left panel) Time evolution of the qubit magnetization $\langle\sigma_z(t)\rangle$ starting from the initial state $|0\rangle$ and evolving toward $|1\rangle$ under the influence of the three different pulse durations.  
(Bottom-left panel) Bloch vector trajectories corresponding to the same pulse durations.  
(Upper-right panel) Gaussian-modulated classical kick pulses for each duration. The associated peak amplitudes are  $\sim$6 mV, $\sim$12 mV, and $\sim$60 mV, respectively.  
(Lower-right panel) Evolution of the $l_1$-norm coherence measure $C_{l_1}$ of the qubit for each pulse case. The pulse area parameter is set to $\alpha = \pi/2$, the drive frequency is $\omega_D = 2\pi \times 4.5~\mathrm{GHz}$, and the qubit dipole moment is taken as $\mu = 3 \times 10^{-25}~\mathrm{C \cdot m}$ for all scenarios.} 
\end{figure*}

The right panel of Fig.~\ref{fig:bloch} presents the Bloch vector trajectory of the qubit initially in the state $|0\rangle$, shown for both $\alpha = \pi$ (top panel) and $\alpha = \pi/2$ (bottom panel). The observed spiral trajectory of the Bloch vector under resonant Gaussian driving reflects the fact that, in the laboratory frame, the qubit’s state precesses at the carrier frequency while undergoing Rabi oscillations. Such spiral patterns are reminiscent of magnetic resonance control in NMR experiments~\cite{ramanathan_nmr_2004}, but here they arise from resonant microwave driving in superconducting qubits.

A comparison of the two trajectories reveals that the driving field implements the $R_x(\alpha)$ gate, where for $\alpha = \pi/2$ the qubit is rotated to the equatorial plane of the Bloch sphere, and for $\alpha = \pi$, the qubit is flipped to the opposite pole, corresponding to a NOT gate. Note that $R_x(\pi) = -iX$, where $X$ is the NOT gate, up to an unobservable global phase. All simulation parameters are listed in the caption of Fig.~\ref{fig:bloch}. These results confirm that under adiabatic drive conditions, where the RWA is valid, the dynamics governed by Eq.~\ref{Eq:Hamilton} effectively implements a generic rotation of the form $R_\nu(\alpha) = \exp\left(-i \frac{\alpha}{2} \sigma_\nu \right)$, where $\nu \in \{x, y, z\}$.

The analysis presented above represents a standard single-qubit gate implementation in the weak-drive regime, as commonly employed in resonant quantum control experiments. While this adiabatic regime is widely used in practice, it serves here primarily as a reference for comparison with our main contribution: the diabatic regime calculations for fast single-qubit gate implementation, which will be summarized below. 

To clearly delineate the diabatic (kick) regime, we consider an ultrashort pulse whose duration $\tau$ is much shorter than the qubit’s intrinsic timescale, i.e., $\omega_0 \tau \ll 1$. In this limit, the control field is modeled by its pulse area rather than a carrier oscillation, and the envelope is effectively an impulse, $E(t)=\alpha\,\delta(t-t_0)$. Accordingly, the drive term reduces to a delta-kick that generates an instantaneous rotation about the $x$-axis. This replaces the carrier-modulated form used in the adiabatic/RWA analysis and leads to the following effective Hamiltonian for the kick regime:
\begin{equation}
\label{eq:H_kick}
H_{\text{kick}}(t)=\frac{\hbar\omega_0}{2}\,\sigma_z\;+\;\hbar\,\alpha\,\delta(t-t_0)\,\sigma_x,
\end{equation}
where $\alpha=\!\int \Omega(t)\,dt$ is the pulse area. The corresponding evolution across the kick is $U=\exp\!\big[-i\,\alpha\,\sigma_x\big]$, highlighting that the diabatic gate is set by the area of the pulse rather than resonance with a carrier frequency. Therefore we will use this model for the diabatic analysis, and reserve the carrier-modulated Hamiltonian in Eq.~\ref{Eq:Hamilton} (RWA regime) solely as a reference for comparison.

To this end, we investigate qubit system dynamics at picosecond pulse durations, maintaining all parameters consistent with the adiabatic regime where the Rotating Wave Approximation (RWA) is applicable. This timescale, being three orders of magnitude shorter than typical realization times for single-qubit quantum logic gates in microwave-based quantum control methods, necessitates exceedingly strong system-field coupling. Figure~\ref{fig:Pico} presents the qubit dynamics for pulse durations of $\tau=0.1$, $0.5$, and $1$ps, respectively. In the upper-left panel, successful population transfer from the initial $|0\rangle$ state to the $|1\rangle$ state is observed across all three scenarios. Our data reveal fidelities of $0.9995$, $0.9992$, and $0.9986$ for each respective duration. These values are nearly commensurate with those achieved for single-qubit logic gate implementations in contemporary superconducting quantum computers using established control techniques. It is crucial to note, however, that this particular implementation transpires over a duration three orders of magnitude shorter than its counterparts. While the deviations in fidelity values across different durations are minor, a trend towards higher fidelities is observed for pulses exhibiting sharper peak profiles.

An examination of the pulse shapes in the upper-right panel of Fig.~\ref{fig:Pico} demonstrates that these picosecond-scale pulses behave as anticipated, exhibiting oscillation-free Gaussian envelopes. Considering the overall application time of $10$ps, it is evident from the figure that only the pulse with the sharpest peak yields the highest fidelity. This observation implies that for successful quantum logic gate realizations in this regime, pulse areas must approximate the Dirac delta function, which represents an ideal kick scenario. Consequently, a practical implication arises: for successful gate operations within multi-pulse sequences under the current parameters, the temporal separation between pulses must be at least two orders of magnitude greater than the individual pulse durations.

Conversely, the lower-left panel of Fig.~\ref{fig:Pico} illustrates the Bloch vector trajectories for the same three distinct pulse durations. Despite remaining in the laboratory frame, a direct continuous rotation is observed, unlike in Fig.~\ref{fig:bloch}, because the kick pulses induce an abrupt, oscillation-free rotation. Although minor deviations in the vector paths are visible, particularly along the $x$-axis, these differences remain on the order of one percent. Notably, the deviation is most pronounced for the longer pulses ($\tau = 0.5$ and $1$~ps), which is consistent with their slightly lower fidelities. These trajectory deviations indicate that the quantum state retains a certain degree of coherence throughout its evolution. Undoubtedly, the $l_1$ norm of coherence, also shown in the lower-right panel, serves as a crucial metric for comprehending pulse efficiencies across all scenarios. As anticipated, coherence is expected to increase during pulse evolution before decaying back to zero; however, varying amounts of non-zero coherence accompany the evolution depending on the pulse durations. Consequently, the minimum coherence generation is observed in the case most closely resembling the ideal Dirac-shaped kick pulse. Collectively, these findings demonstrate that single-qubit logic gates in the diabatic regime can be implemented with fidelities comparable to their RWA counterparts through meticulous analysis.

Beyond the numerical analyses presented above, another crucial aspect to highlight is our consistent use of $\alpha=\pi/2$ for the implementation of the NOT gate in the diabatic calculations. While the parameter $\alpha$ is still defined as the time integral of the instantaneous Rabi frequency, the pulse area theorem, which is valid within the Rotating Wave Approximation, no longer holds in the diabatic regime. In this limit, the rotation angle is not solely determined by $\alpha$, as counter-rotating terms and the breakdown of the RWA modify the qubit's response to the drive. The key phenomenon here is that the contribution from the counter-rotating term constructively combines with that of the rotating term, thereby effectively doubling the driving interaction.

Our study is centered on investigating single-qubit logic gates in two-level systems within the diabatic regime, employing extremely short pulse areas with realistic parameters. For this purpose, our calculations utilized parameters characteristic of transmon qubits, which represent the most prevalent qubit realization architecture. Transmon qubits, as a low-anharmonicity superconducting qubit architecture, are distinguished by their extended coherence times and well-characterized energy levels under microwave control. By design, the level structure of this architecture is weakly anharmonic, which inherently poses potential challenges when subjected to high-speed and powerful (diabatic) drives. In transmon qubits, where the objective is to achieve an effectively two-level quantum system, the energy shift to the third energy level is typically around $200-300 \text{ MHz}$~\cite{koch_charge-insensitive_2007, krantz_quantum_2019}. Indeed, in our work, picosecond-scale pulses with amplitudes on the order of tens of millivolts could induce undesirable population transfer (leakage) to higher energy levels~\cite{motzoi_simple_2009}, rendering the transmon architecture a less favorable choice for such strong pulse fields.

Conversely, the Flux qubit architecture~\cite{chang_reproducibility_2022,hita-perez_three-josephson_2021}, which offers greater anharmonicity under strong control pulses due to its double-tunneling barrier and larger level separations, might emerge as a more suitable alternative for kick pulses. Consequently, implementing logic gates with ultra-short pulse areas in the diabatic regime presents specific limitations when applied to transmon qubits.
\section{CONCLUSIONS}

In this study, the dynamics of a two-level quantum system subjected to a classical driving field have been thoroughly investigated under two distinct control regimes: adiabatic (Rotating Wave Approximation (RWA) regime) and diabatic (kick pulse regime). Particular emphasis was placed on realizing a single-qubit NOT gate, analyzing quantum state transitions induced by Gaussian-modulated pulses of varying durations and amplitudes. Simulations conducted in the adiabatic regime demonstrated the expected classical Rabi oscillations and Bloch sphere rotation dynamics with $\pi$ and $\pi/2$ pulses. Concurrently, using identical parameters, the system proved capable of achieving NOT gates with fidelities exceeding $99.9\%$ under short, powerful pulses on the picosecond scale. 

Furthermore, the instantaneous coherence generated during these brief pulses was analyzed as a function of the applied pulse duration and sharpness. The results indicated that the shortest and highest-amplitude pulses, representing pulse modulations closest to the ideal kick shape, are crucial for approaching optimal logic gate fidelity in the diabatic regime.

The reported findings carry significant implications for the potential reduction of quantum logic gate realization times by orders of magnitude in current noisy intermediate-scale quantum computers. Although our computations in this study were predicated on the transmon qubit architecture, we believe that the inherent limitations of transmon qubits when faced with short and strong fields will inspire experimentalists to explore alternative architectural solutions beyond this paradigm.

\begin{acknowledgments}
D.~T. acknowledges support from the Scientific and Technological Research Council of Turkey (T\"{U}B\.{I}TAK)-Grant No. 124F472. S.~\c{C}. acknowledges support from the Scientific and Technological Research Council of Turkey (T\"{U}B\.{I}TAK)-Grant No. 122F298. The authors also wish to thank Ferdi Altintas for his valuable contributions to the development of the main problem framework and for the insightful discussions. 

\end{acknowledgments}

\appendix
\section{Pulse Amplitude}
\label{appendix:pulse}

To characterize the overall pulse strength, we introduce a dimensionless parameter $\alpha$, defined as the time integral of the Rabi frequency:
\begin{align}
\alpha = \int_{-\infty}^{\infty} \Omega(t) \, dt 
       &= \frac{\mu A_0}{\hbar} \int_{-\infty}^{\infty} 
          \exp\left(-\frac{(t - t_0)^2}{\tau^2}\right) dt \notag \\
       &= \frac{\mu A_0}{\hbar} \sqrt{\pi} \tau.
\end{align}
This implies that the peak electric field amplitude is related to $\alpha$ by $A_0 = \frac{\alpha\hbar}{\mu\sqrt{\pi}\tau}$.


\section{Voltage Interpretation and Physical Units}
\label{appendix:Voltage}

The electric field required for a $\pi$-pulse must be generated via a voltage drive. In experimental settings, this voltage is typically applied across a transmission line or qubit structure, and the electric field $E(t)$ is directly related to the applied voltage $V(t)$ by:
\[
E(t) = \frac{V(t)}{L_{\text{eff}}},
\]
where $L_{\text{eff}}$ is an effective length scale determined by the geometry and placement of the qubit and the control line. For planar transmon devices, $L_{\text{eff}}$ is typically on the order of tens of micrometers. In this work, we use $L_{\text{eff}} = 20~\mu\mathrm{m}$.

Substituting this into the Rabi condition, the Hamiltonian coupling can be written as:
\[
H_{\text{drive}}(t) = \mu E(t) \sigma_x = \mu \frac{V(t)}{L_{\text{eff}}} \sigma_x.
\]
For a Gaussian pulse envelope that achieves a $\pi$-rotation (a NOT gate), the required peak voltage $V_{\text{peak}}$ is derived from the integral of the Rabi frequency. For a drive with amplitude $E_{\text{peak}}$ and Gaussian width $\tau$, the integral is $\int \Omega(t) dt = \frac{\mu E_{\text{peak}}}{\hbar} \sqrt{\pi}\tau = \pi$. Substituting $E_{\text{peak}} = V_{\text{peak}} / L_{\text{eff}}$, the required peak voltage for a $\pi$-pulse is given by:
\[
V_{\text{peak}} = \frac{\pi \hbar L_{\text{eff}}}{\mu \sqrt{\pi} \tau} = \frac{\sqrt{\pi} \hbar L_{\text{eff}}}{\mu \tau}.
\]

Using realistic values for a transmon qubit: effective dipole moment $\mu = 3 \times 10^{-25}~\mathrm{C \cdot m}$, pulse width $\tau = 23~\mathrm{ns}$, and effective length $L_{\text{eff}} = 20~\mu\mathrm{m}$, we find:
\begin{align*}
V_{\text{peak}} &= \frac{\sqrt{\pi} \cdot (1.054 \times 10^{-34}~\mathrm{J \cdot s}) \cdot (20 \times 10^{-6}~\mathrm{m})}{(3 \times 10^{-25}~\mathrm{C \cdot m}) \cdot (23 \times 10^{-9}~\mathrm{s})} \\
&= \frac{1.772 \times 10^{-34}~\mathrm{J \cdot s} \cdot 20 \times 10^{-6}~\mathrm{m}}{6.9 \times 10^{-33}~\mathrm{C \cdot m \cdot s}} \\
&\approx 5.42 \times 10^{-5}~\mathrm{V} \approx 54.2~\mathrm{\mu V}.
\end{align*}
This value is well aligned with the experimentally observed voltage ranges for single-qubit control in superconducting circuits, which typically span hundreds of nanovolts to tens of millivolts.

\bibliographystyle{apsrev4-2}
\bibliography{Refsbib}

\end{document}